\documentclass[iop]{emulateapj}

\newcommand{\htwo}{H$_2$}
\newcommand{\av}{$A_V$}

\newcommand{\ci}{\ion{C}{1}}
\newcommand{\cii}{\ion{C}{2}}
\newcommand{\fn}{$f^n$}
\newcommand{\fN}{$f^N$}

\shortauthors{BURGH, FRANCE, \& JENKINS}
\shorttitle{CARBON CONTENT IN THE TRANSLUCENT ISM}

\begin{document}

\title{Atomic and Molecular Carbon as a Tracer of Translucent Clouds}

\author{Eric B. Burgh, Kevin France}
\affil{Center for Astrophysics and Space Astronomy, University of Colorado
  - Boulder, 593 UCB, Boulder, CO 80309}
\and
\author{Edward B. Jenkins}
\affil{Princeton University Observatory, Peyton Hall, Ivy Lane, Princeton, NJ 08544.}


\begin{abstract}
Using archival, high-resolution far-ultraviolet \textit{HST}/STIS spectra
of 34 Galactic O and B stars, we measure \ion{C}{1} column densities and
compare them with measurements from the literature of CO and \htwo\ with
regard to understanding the presence of translucent clouds along the
line-of-sight.  We find that the CO/\htwo\ and CO/\ci\ ratios provide good
discriminators for the presence of translucent material, and both increase
as a function of molecular fraction, $f^N = 2N(H_2)/N(H)$.  We suggest that
sightlines with values below CO/\htwo\ $\approx10^{-6}$ and CO/\ci\
$\approx1$ contain mostly diffuse molecular clouds, while those with values
above sample clouds in the transition region between diffuse and dark.
These discriminating values are also consistent with the change in slope of
the CO v. \htwo\ correlation near the column density at which CO shielding
becomes important, as evidenced by the change in photochemistry regime
studied by Sheffer et al. (2008).  Based on the lack of correlation of the
presence of translucent material with traditional measures of extinction we
recommend defining 'translucent clouds' based on the molecular content
rather than line-of-sight extinction properties.
\end{abstract}

\section{Introduction}

In the traditional three phase interstellar medium (ISM), the cold neutral
medium (CNM) contains the bulk of the interstellar ``clouds''.  These
clouds range from the dense dark clouds, the centers of which are the
birthplaces of the next generation of stars, typically observed by CO radio
emission, to the tenuous diffuse clouds, which can be probed by absorption
spectroscopy of background stars.

In the literature, these are typically defined by their extinction
properties, with the diffuse ISM having \av$\lesssim1$ and the dense clouds
having \av$\gtrsim5$ \citep{vDB88}.  This distinction by extinction
indicates the importance of the ultraviolet portion of the interstellar
radiation field on the photochemistry in the clouds.  The diffuse ISM
is fully exposed, and therefore most species exist in atomic or ionized
forms.  In dense clouds the ultraviolet light is sufficiently shielded such
that molecules are not photodissociated and virtually all the hydrogen
exists in molecular form (\htwo), with carbon monoxide (CO) the second most
abundant molecule.

Sightlines with intermediate extinction tend to be referred to as
``translucent''.  A translucent sightline may simply be the concatenation
of multiple diffuse clouds, resulting in a higher value of \av\ than is
typical of a sightline that might pass through a single diffuse cloud.  One
may also sample the transition region between the diffuse and dense clouds,
where species make the transition between their atomic and molecular forms
-- what could be considered a translucent ``cloud''.  In these regions the
chemistry is sensitive to the physical conditions, such as density,
temperature and radiation field, which may be changing rapidly from one
location to the next.  With increased depth into a cloud comes an increased
fraction of molecular content, both in \htwo\, because of self-shielding,
and CO, because of shielding by \htwo, as well as the attenuation of
photodissociating flux by dust \citep{vDB88}.  In the case of hydrogen, the
transition from atomic to molecular form is quite sharp, with the local
molecular fraction ($f^n = 2n(H_2)/n_H$, $n\equiv$ space density) reaching
essentially a value of one at very short depths into a cloud.  Observations
of the integrated molecular fraction ($f^N = 2N(H_2)/N(H)$, $N\equiv$
column density) show a sharp jump from values less than $\sim10^{-4}$ to
values larger than $\sim0.01$ consistent with the presence of this
transition \citep{Spitzer75,Gillmon06}, and as a consequence \htwo\ is
ubiquitous in the diffuse ISM \citep{Shull00}. The gas-phase carbon makes a
transition from ionized form (\cii) to molecular (CO).  Although the
calculated depth at which this transition happens and the relative
abundances of all the carbon-bearing species depends on the physical parameters
and chemical networks used, this qualitative description appears in all
models \citep{Rollig07}.

The far-ultraviolet (far-UV: $\lambda\lesssim2000$\AA) portion of the
electromagnetic spectrum allows for the direct observation of both \htwo\
and CO in the ISM through absorption spectroscopy towards distant O and B
stars \citep{Savage77,Rachford02,Federman80,Burgh07,Sheffer08}.  In
particular, the study of \citet{Rachford02} investigates the molecular
content and physical properties along lines-of-sight considered translucent
by their extinction properties, reaching the conclusion that they are
likely the projections of multiple diffuse clouds with low extinction
rather than a single translucent cloud with higher extinction.
\citet{Burgh07} suggested that the ratio of CO to \htwo\ would serve as a
better discriminator of diffuse and translucent clouds and they  measure a
transition from low to high values of CO/\htwo\ with increased molecular
fraction.  The CO abundance is more sensitive to the effects of geometry,
dust shielding and fragmentation of clouds, and thus if a given sightline
were simply a collection of diffuse clouds, there would not be an increase
in the measured CO/\htwo\ \citep{Kopp00}.  If the term ``translucent''
should refer to the transition region between the physical states of
``diffuse'' and ``dense'' then these results demonstrate a weakness in
defining translucent clouds based solely on line-of-sight extinction or
\htwo\ properties.

\citet{Snow06} recommend a different definition for ``translucent'', based
upon the carbon content, specifically the transition from ionized to
molecular form.  This suggests that observations of neutral carbon
(\ci), also available in the far-UV \citep{Jenkins79,Jenkins01}, may
give further insight as to the presence of translucent clouds along a given
sightline.  Steady-state PDR models predict that \ci\ is formed through
recombination of \cii\ and photodissociation of CO, and its abundance
should peak in the region between the ionic and molecular parts of the
cloud.  \citet{Snow06} then further break down cloud structure into the
following categories (see their Table 1): diffuse atomic, where the low
molecular fractions mentioned earlier are observed; diffuse molecular,
where the hydrogen is primarily molecular, but carbon still in ionized
form; translucent, where the carbon makes the transition to molecular; and
dense, where both the hydrogen and carbon are fully molecular.  We agree
with this type of categorization, eschewing the use of extinction as the
sole discriminator.

In this work, we expand upon the study of \citet{Burgh07} by including
measurements of the \ci\ lines observed in the \textit{HST}/STIS E140H
mode.  We compare these measurements to CO and \htwo\ in an attempt to
better isolate the transition from diffuse to translucent and better
understand the conditions under which this transition occurs.

\section{Column Density Determinations}

\subsection{\htwo\ and CO}

This work expands on the data set used in \citet{Burgh07}, by adding
sightlines with \htwo\ column density determinations by \citet{Sheffer08},
\citet{Nehme08a}, \citet{Rachford09}, and \citet{Shull09}.  CO column
density determinations are from \citet{Burgh07}, \citet{Sheffer08}, as well
as six new fits, determined in the same manner as in \citet{Burgh07}.
Although a larger dataset exists with both \htwo\ and CO column densities,
we are presenting here only those for which there existed adequate
wavelength coverage of STIS E140H data for the \ci\ measurement, described
in the next section.  

The total number of sightlines is 34 and range from diffuse to translucent
material, with extinctions from E(B-V) = 0.07 -- 0.61, Log N(\htwo) = 18.73
-- 20.92, with molecular fractions ranging from 0.02 -- 0.76.  The CO
column densities have some upper limits around 12.3 but mostly detections
in the range Log N(CO) = 12.98 -- 16.13.  Table 1 lists
the column densities for the hydrogen bearing species and Table
2 lists those for carbon, both with appropriate
references.

\subsection{\ion{H}{1}}
\label{hone}

For the most part, the \ion{H}{1} column densities are taken from the
literature and determined from profile fits of the Ly-$\alpha$ line
\citep{Diplas94,FM90,Rachford02,Cartledge04,Jensen07}.  However, in a few
cases, the spectral type of the star is late enough such that contamination
from the stellar Ly-$\alpha$ prevents an accurate determination of the
interstellar atomic hydrogen column density.  For three cases (HD~27778,
HD~147888, and HD~203532), we use the \citet{Bohlin78} relationship between
$N(H_\mathrm{tot}$) and $E(B-V)$ to determine the total hydrogen column
density.

HD~102065, however, poses an interesting problem.  \citet{Burgh07} state
that the presence of weak \ion{C}{4} absorption features seemed to indicate
that the original spectral type identification of B9IV was wrong and
reclassified the star as a B2V, an early enough type to allow for
Ly-$\alpha$ profile fitting.  However, the analysis of \citet{Nehme08a}
suggests that the \ion{C}{4} could be from hot interstellar gas along the
line-of-sight, and thus the original identification may be
correct.  For this study, we will assume that is true and determine the
\ion{H}{1} column from the Bohlin relationship.  This is not an
insignificant issue because the difference in molecular fraction between
the two cases is large (\fN=0.10 if B2V and \fN=0.69 if B9IV).  This will
be commented upon later in the results section.

\subsection{\ci}

The \ci\ column densities were determined following the method of
\citet{Jenkins01}, which takes advantage of a large number of multiplets
including a wide range in line strengths to provide solutions for the
apparent column densities $N_a$ as a function of radial velocity.  This is
done for each of the \ci,\, \ci$^*$, and \ci$^{**}$ levels.  The column
density for each level is obtained by integrating over the full velocity
range\footnote{It is of interest to note that some sightlines show velocity
components with more column in the excited states than the ground states,
indicative of higher thermal pressure.  However, these are typically seen
at higher velocities, thus unrelated to the clouds in which the molecules
reside, and also amount to a small fraction of the total column.  We have
made no effort to isolate any specific velocity range for integration in
order to exclude these regions.}  and then the total column density,
presented in Table 2, is the sum of each level.  The
\citet{Jenkins01} method is not conducive to determining a random error for
the velocity integrated column density, so we adopt an error of 0.05 dex,
or $\sim12\%$, which is consistent with similar determinations from other
methods.

The $N_a$ profiles were determined using two values of the ``transition
intensity'' mentioned in Equation (7) in Section 5.2.1 of
\citet{Jenkins01}.  This parameter reduces the weight on the parts of the
lines that are strongly saturated in the original data.  In 31 of the 34
cases, the lower ($I_t=0.2$) and higher ($I_t=0.4$) values produced column
densities with differences less than 0.05 dex.  Furthermore, for the
sightlines that will be defined as ``diffuse'' by the criteria mentioned
later, the maximum difference is 0.02 dex, which is within our adopted
minimum error in N(\ci).  In the three remaining cases the differences
were: 0.1 dex (HD~24534), 0.16 dex (HD~147888), and 0.24 dex (HD~203532).
This indicates that in these cases, there is likely to be unresolved
saturated velocity structure affecting the result, and we assign a larger
error (0.10 dex) to these.  To err on the side of being conservative, we
have used the higher value of $I_t$ for all cases because it is likely to
produce a more accurate result for those sightlines that suffer from
unresolved saturated velocity structure and makes little difference for
those that do not.

It is important to note here that we are using the oscillator strengths
from \citet{Jenkins01}, which differ from those of theoretical and
experimental derivations compiled in \citet{Morton03}.  This
can result in large differences in derived column density; e.g.,
\citet{Sonnentrucker02} and \citet{Sonnentrucker03} derive total \ci\
column densities of Log N(\ci)=15.37 and Log N(\ci)=15.57 for HD~192639 and
HD~185418, respectively.  These differ by factors of about 4 and 7 from the
values determined in this study.  The differences in oscillator strengths
grow systematically with a decrease in strength, as shown in Figure 3 of
\citet{Jenkins01}, so the greater the column density the larger the
deviation because more weight will be placed on the weaker lines in the
derivation.  We believe that given the self-consistency in the results
provided by the rederived $f$-values we are justified in using them.
Furthermore, the trends found in our study should be unaffected by these
kinds of systematic changes in column density, except for potential slope
changes in correlations because of the strength dependence of the deviations.

\section{Results}

\begin{figure}
\epsscale{0.9}
\plotone{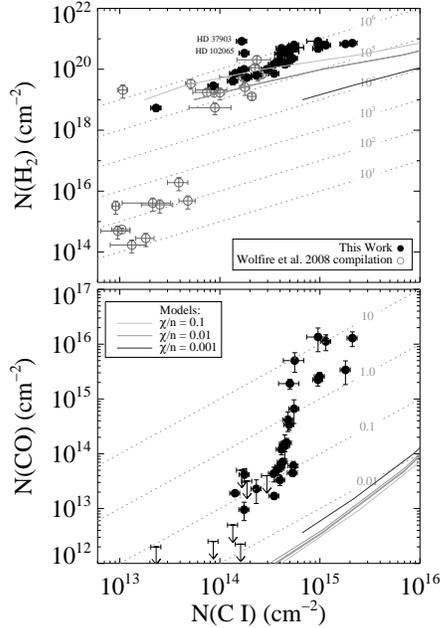}
\caption{\htwo\ and CO versus \ci. Arrows designate upper limits on the CO
  measurement.  Dashed lines show the ratio of the each species with \ci.
  Overplotted are models from \citet{Visser09} for varying values of the
  ratio of the strength of the interstellar radiation field to the density
  ($\chi/n$).
\label{covci}}
\end{figure}

Figure \ref{covci} shows the relationships between \htwo, CO and \ci.
Included in the upper panel are \ci\ values from the literature and
compiled in \citet{Wolfire08} to give a broader context to the data
presented here.  The lack of sightlines with \htwo\ column densities in the
range of about $10^{17}$ to $10^{18}$ represents the turn-on of \htwo\ due
to self-shielding \citep{Gillmon06}, as evidenced by the fact that the low
\htwo\ column sightlines all have a \fN$\lesssim10^{-4}$ and the sightlines
in our sample all have \fN$>0.01$.  In the lower panel, the CO/\ci\ shows a
trend with a steep slope, $N(\mathrm{CO})\propto N($\ci$)^{2.5}$, with the
CO column density rising over two orders of magnitude for a small range in
\ci. There may be a hint of a slope change similar to that found in the CO
versus \htwo\ relationship near $N(\mathrm{CO})$ of $10^{14}$
\citep{Sheffer08}.  At about Log N=14.7 there is as much CO as \ci\ along
the line-of-sight, and the trend suggests the observation of the transition
from diffuse to translucent molecular material, where CO becomes the
dominant carbon-bearing species.

\begin{figure*}
\plotone{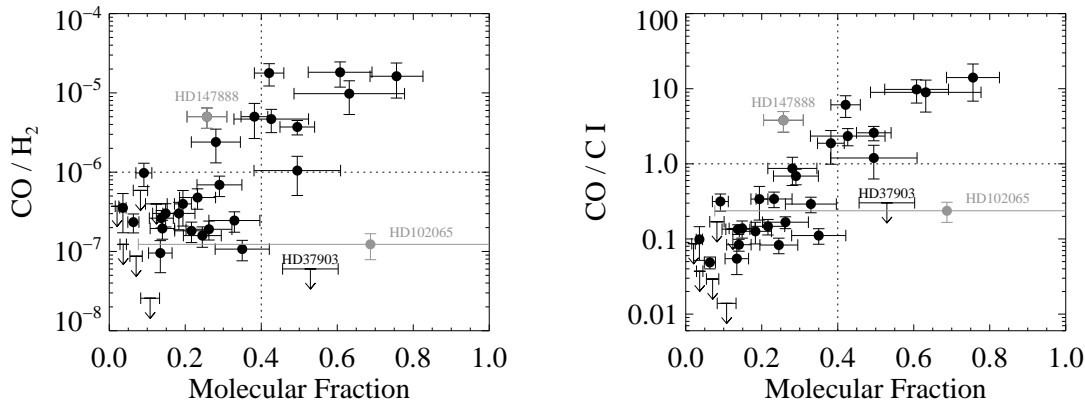}
\caption{CO/\htwo\ (left) and CO/\ci\ (right) versus molecular fraction.
  The CO/\htwo\ transitions from low (diffuse) to high (translucent)
  between \fN =0.2--0.4, with all values above $10^{-6}$ by \fN=0.4.  The
  CO/\ci\ shows a linear relationship with \fN, with all values above 1 by
  \fN =0.4.  HD102065 (the \fN\ error for which is quite large due to
  uncertainties in the spectral classification), HD147888 and HD37903 are
  notable exceptions and are discussed in the text.
\label{versusmf}}
\end{figure*}

The CO/\ci\ is similar to CO/\htwo\ in its utility as a discriminator
between the diffuse and translucent regimes.  Figure \ref{versusmf} shows
both ratios versus molecular fraction.  As discussed by \citet{Burgh07},
the CO/\htwo\ shows a transition from low to high between \fN = 0.2 and
\fN = 0.4.  CO/\ci\ shows a stronger correlation ($r \sim 0.8$, excluding the
outliers) with values above 1 for \fN$>0.4$.

There are a few notable exceptions to the relationships observed in Figure
\ref{versusmf}: HD~147888, HD~37903, and HD~102065.  HD~147888 ($\rho$ Oph
D) lies in a complex area of the sky, deeply investigated by
\citet{Snow08}.  There is no direct \ion{H}{1} column density
determination, so it is possible that the molecular fraction is
underestimated.  In terms of molecular content it appears more like a
translucent sightline than other sightlines of similar \fN.  As mentioned
in Section \ref{hone}, the \fN\ for HD~102065 depends strongly on the
spectral type classification of the star, which might be in doubt because
of observed \ion{C}{4} absorptions, which may or may not be of stellar
origin.  The molecular content is consistent with a diffuse cloud, as would
be expected if the lower \fN\ were correct, and \citet{Nehme08b} do find a
best model fit for low density ($n \sim 80$~cm$^{-3}$).  The HD~37903
sightline is unique in our sample for having its \htwo\ in very close
proximity to the star, such that there are observed absorptions out of
vibrationally excited levels.  \citet{Meyer01} show that this is consistent
with a dense parcel of gas being illuminated by a flux about 1000 times the
strength of the average ultraviolet interstellar radiation field.  This
intense flux would be strong enough to photodissociate the CO in that
cloud, and thus, even though the molecular fraction is about 0.5, there is
at best only a marginal detection of CO.  It would make for an interesting
comparison to explore, with a rejuvenated STIS, other nebular environments,
such as the sightlines to HD~34078, which also shows absorption from highly
excited \htwo\ \citep{Boisse05}, and HD~200775, where \textit{IUE} data
show a very high CO column density \citep{Knauth01}.

\begin{figure}[b]
\epsscale{1.0}
\plotone{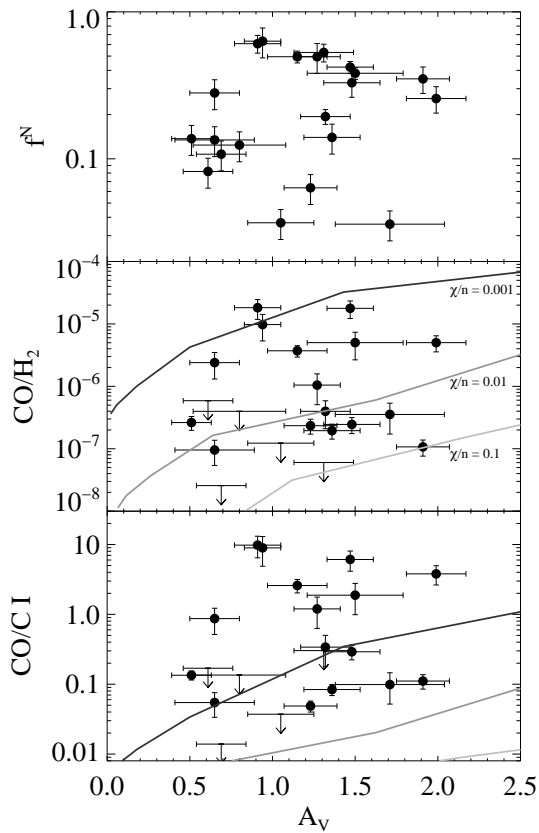}
\caption{Molecular fraction, CO/\htwo\ and CO/\ci\ versus \av, for a
  sub-sample of 21 stars with \av\ determined by \citet{Valencic04}.
  Overplotted are models from \citet{Visser09} for varying values of the
  ratio of the strength of the interstellar radiation field to the
  density ($\chi/n$).  These show that translucent clouds could be present 
  along sightlines with low \av\ if $\chi/n$ is also low.
\label{versusav}}
\end{figure}

Figure \ref{versusav} shows, for a sub-sample of 21 sightlines with \av\
determined by \citet{Valencic04}, scatter plots of \fN, CO/\htwo\ and
CO/\ci\ against \av.  No significant correlations or trends are seen,
indicating that line-of-sight extinction is not a good measure of the
molecular content in the transition between diffuse and translucent clouds.
In particular, one should note that some of the largest CO column densities
are seen for sightlines with \av~$\lesssim1$.  Translucent cloud models are
overplotted and will be discussed in Section \ref{models}.

\section{Discussion}

\subsection{Diffuse to Translucent Transition}

The measured quantities along a given sightline are, by nature, integrated
quantities and may not be indicative of the local values of the material
through which the line-of-sight passes.  Historically, the word
``translucent'' has been applied to sightlines of a given \av; however,
following \citet{Snow06} we believe the distinction between diffuse and
translucent should be made based on the local values of the material
sampled.  Therefore, we use the words diffuse and translucent to describe
sightlines based on the type of cloud sampled (i.e., low or high local
neutral carbon content), independently of the line-of-sight extinction.

\begin{figure}
\plotone{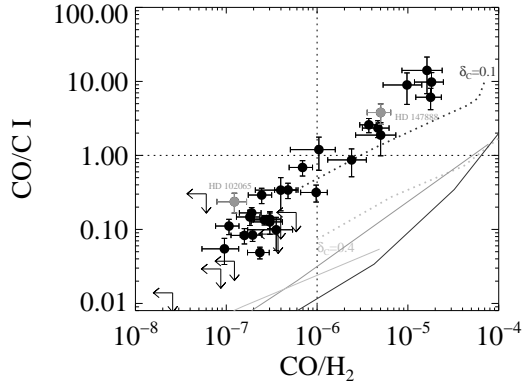}
\caption{CO/\ci\ versus CO/\htwo. Arrows designate upper limits on the CO
  measurement.  We will identify those sightlines with CO/\htwo$\gtrsim
  10^{-6}$ and CO/\ci$\gtrsim 1$ as ``translucent''.  The solid lines are
  models from \citet{Visser09} with colors following those in the previous
  figures.  The dashed lines are translucent cloud models from
  \citet{vDB88} with two different values of carbon depletion.
\label{cocivcoh2}}
\end{figure}

Based on the correlations seen in Figure \ref{versusmf} we believe we are
seeing the transition from diffuse to translucent clouds being probed for
sightlines with \fN$\gtrsim0.4$.  More specifically we suggest the values
of CO/\htwo~=~$10^{-6}$ and CO/\ci~=~1 as discriminating values between
sightlines with diffuse and translucent clouds.  Those sightlines with
measurements above these values are almost completely inclusive of each
other, independently of molecular fraction.  This is shown explicitly in
Figure \ref{cocivcoh2}.  Although we expect \ci\ to peak in the translucent
cloud regime, it is primarily the steep increase in CO abundance that is
driving the observations.  The standouts from Figure \ref{versusmf} are
well-behaved in Figure \ref{cocivcoh2}, indicating that these ratios are
more reliable than molecular fraction alone in distinguishing between
sightlines with diffuse and translucent clouds; e.g., regardless of the
uncertainty in the \ion{H}{1} determinations for HD~102065 and HD~147888,
these sightlines clearly have diffuse and translucent material, respectively.

This is supported by the work of \citet{Sheffer08}, who recently analyzed a
large sample of Galactic sightlines with far-UV data of CO and \htwo, in
addition to optical data of other carbon-bearing molecules. They conclude
that at higher column densities there is a distinct change from low- to
high-density photochemistry.  They delineate the change at Log N(CO)=14.1
and Log N(\htwo)=20.4, the point at which the CO vs. \htwo\ correlation
changes slope.  Above Log N=20.4 in \htwo\ there is still a wide range in
CO/\htwo\ and CO/\ci, most likely due to sightline to sightline variations
in the ratio of the strength of the radiation field to the density
($\chi/n$ hereafter) and/or cloud structure/fragmentation.  Above Log
N=14.1 of CO we find some sightlines with total neutral content (see
discussion below) that look diffuse.  If we make a cut based on the total
neutral carbon content we find that Log N(CO)=14.6 is a better value for
discriminating the two regimes and is fully consistent with the values we
suggested above of CO/\htwo~=~$10^{-6}$ and CO/\ci~=~1.

\subsection{Single versus Multiple Clouds}

In the range of \av\ sampled in this study, extinction properties give no
indication of the physical or chemical conditions in the clouds sampled
along the line-of-sight; therefore, a sightline with a given \av\ may
sample a thick cloud with translucent material, or simply be overlapping
clouds with more diffuse material.

This is an important distinction to make, as evidenced by the studies of
translucent sightlines by Rachford et al.(2002, 2009).  They came to the
conclusion that their lines of sight sample multiple diffuse clouds rather
than truly ``translucent'' clouds; however, we note that 31/38 have
\fN~$>0.2$ and about half have \fN~$>0.4$.  This suggests that, despite
their expectations for high extinction and \fN~=~1, it is likely that a large
fraction of their sightlines do include clouds with significant
molecular content. Indeed, they admit that were they to relax their
requirement for large \av\ they would conclude that ten of their
lines-of-sight contain translucent clouds.  These ten, which include
HD~24534 and HD~27778 (two of our highest CO/\htwo\ values), were noted
specifically for having larger \fN\ and lower \htwo\ rotational temperatures.

Mathematically speaking, \fN\ is an average value for the sightline, which
must sample some material with \fn~$>$~\fN.  The presence of a translucent
cloud, where \fn~$\approx$~1, could be somewhat hidden by an extensive
foreground of diffuse atomic material, where \fn\ is likely to be less than
$10^{-4}$.  \citet{Browning03} claim that, based on their \htwo\ properties
alone, they were unable to model these sightlines as a single cloud of high
extinction, but instead they were more successful with concatenations of
smaller clouds of lower extinction each; i.e., there must be multiple
pathways for radiation getting in.  However, this is in contradiction with
the high abundance of CO, as shown in the CO/\htwo\ and CO/\ci\
relationships.

We note that two of the stars in that sample, HD~192639 and HD~185418, were
studied in detail by \citet{Sonnentrucker02} and \citet{Sonnentrucker03},
respectively.  In both cases, they came to the conclusion that translucent
material was not seen, which would seem to bolster the Rachford et
al. claim.  However, these prove to be good examples in our current study:
HD~192639, for example, has a high \av\, (1.91) and reasonably high
molecular fraction (\fN~=~0.35), but has both a low CO/\htwo\
($1.07\times10^{-7}$) and CO/\ci\ (0.11).  Both of these values are
consistent with a diffuse cloud.  HD~185418 is a borderline case
(CO/\htwo~$= 1.05\times10^{-6}$ and CO/\ci\ = 1.25).

According to models of individual clouds, the transition from atomic to
molecular hydrogen, dominated by self-shielding, happens very quickly, over
a short range of extinction.  Therefore, the transition from \cii\ to CO is
occurring in a region where $f^n\approx1$, regardless of the value of \fN.
Outside of the diffuse molecular clouds, \fN\ is typically $<10^{-4}$ and
probably consistent with \fn.  Thus, for $f^N\gtrsim0.01$, the value of
\fN\ is indicative of the amount of hydrogen along the line of sight that
has $f^n\approx1$ and the trends shown in Figure \ref{versusmf} put a
constraint on the fragmented structure of the diffuse ISM.  With increasing
molecular fraction, we observe an increasing CO content such that the high
\fN\ sightlines cannot be too porous, letting in CO-photodissociating
far-UV radiation, but must include significant amounts of translucent
material.

\subsection{Neutral Carbon Fraction}

\begin{figure}[b]
\plotone{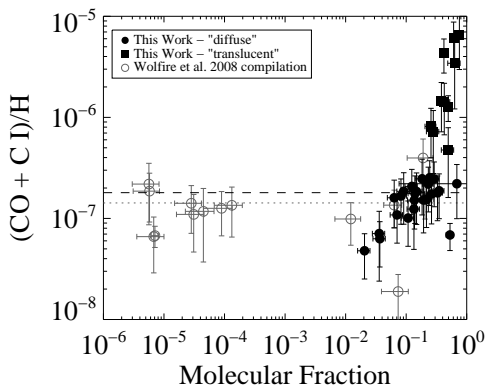}
\caption{Neutral carbon content (\ci\ + CO) per hydrogen nucleon as a
  function of molecular fraction.  Filled squares and circles are for those
  sightlines sampling translucent and diffuse material respectively, while
  open circles represent the sample from \citet{Wolfire08} for Log
  N(\ion{C}{1}) $<14.0$ -- for these, we expect the CO contribution to be
  negligible.  Dashed and dotted lines are the average values for the
  diffuse sightlines from this work and the Wolfire compilation
  respectively, excluding the outliers with (\ci\ +
  CO)/H $<8\times10^{-8}$, which are mostly from sightlines that pass through
lower density material and/or higher than average radiation environments,
such as HD~37903, mentioned previously.  HD~93222 and HD~93843 reside near
the Carina nebula, while HD~121968 is a halo star, lying behind the Radio
Loop I and IV SNRs \citep{Sembach97}.  The lowest neutral fraction is
towards HD~143018 ($\pi$ Sco).
\label{cneuth}}
\end{figure}

It is possible to put the more unusual standouts with respect to molecular
fraction in context by looking at the overall neutral fraction of the
carbon -- the total neutral carbon content per hydrogen nucleon, (\ci\
+ CO)/H.  This is shown as a function of molecular fraction in Figure
\ref{cneuth}. Plotted with our results are data from the literature as
compiled by \citet{Wolfire08} to demonstrate the full run over the widest
range, including sightlines sampling what is most likely primarily
``diffuse atomic'' material.  These additional points have no CO
measurements; however, we use only those sightlines for which
N(\ci)$<14.0$, where we expect the CO contribution to be negligible.  The
gap centered around \fN~=~$10^{-3}$ separates the atomic from the molecular
clouds \citep{Spitzer75,Gillmon06}.

In this representation it is clear how the diffuse and translucent
sightlines differ.  Also, it shows how the diffuse sightlines in our sample
have a consistent ionization state with those primarily atomic sightlines,
with very small molecular fraction, suggesting that the \ci\ observed along
these sightlines most likely exists in the diffuse atomic material, where
the \ci\ abundance will be governed by a balance between photoionization
and recombination.  Those sightlines with higher molecular fraction but
lower neutral carbon abundance (such as HD~102065) may indeed, as suggested
by \citet{Rachford02}, be a concatenation of diffuse clouds or be sampling
material with higher ratios of radiation intensity to density.

\begin{figure}
\plotone{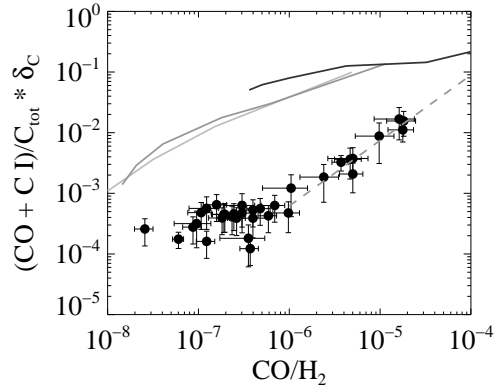}
\caption{Fraction of gas-phase carbon in neutral form multiplied by the
  depletion factor $\delta_\mathrm{C}$ as a function of CO/\htwo.  The red
  line is a fit to the linear rise in the translucent regime (CO/\htwo\
  $>10^{-6}$).  Solid lines are models from \citet{Visser09} with colors
  following those in the previous figures.
\label{neutfrac}}
\end{figure}

The overall fraction of carbon that is in neutral form rises sharply in the
translucent regime.  Figure \ref{neutfrac} shows this as a function of
CO/\htwo, plotting the value (\ci + CO)/H $\div$ (C/H)$_\mathrm{cosmic}
\times \delta_\mathrm{C}$.  We use the \citet{Holweger01} solar abundance
value (391 ppm) for the cosmic carbon abundance and leave the depletion
factor as a free parameter.  The dashed line is a fit to the rise for
values of CO/\htwo\ greater than $10^{-6}$, i.e., the translucent cloud
regime, and produces a power law slope of essentially unity:
$\mathrm{Log}_{10}(\mathrm{C_{neutral}/C_{total}})\propto
\mathrm{Log}_{10}(\mathrm{CO}/\mathrm{H}_2)^{1.08\pm0.23}$.  For a constant
depletion factor, the values in the plot will shift up with the slope
unchanged, e.g., if $\delta_\mathrm{C}=0.1$ then our most translucent
sightlines reach nearly 17\% neutral fraction for the integrated column
densities.  However, if there is more depletion in translucent clouds than
in the diffuse ISM, the slope may be increased.

\subsection{Comparison to Translucent Cloud Models\label{models}}

Because we only measure integrated/averaged quantities we need to make
comparisons to models of the ISM in order to get an understanding of what
the local values are along a line-of-sight.  Figures 1, 4 and 6 show models
from the recent work of \citet{Visser09}, who revisited the translucent
cloud models of \citet{vDB88}, using updated values of oscillator
strengths, expanding the range of column densities (which include the
suprathermal chemistry regime suggested by \citet{Sheffer08}), and modeling
more isotopologues of CO.  In particular, we show the curves for three
values of the ratio of the strength of the interstellar radiation field
($\chi$ in units of the \citet{Draine78} field) to the gas density ($n$).
These models do a good job at matching observations of \htwo\ and CO (see
Visser's Figure 7); however, we are finding a large discrepancy between the
models and the data for \ci.

There is even poorer agreement with the data for the models from
\citet{Rollig07}, which are specifically for warm and dense
photo-dissociation regions and we believe are not well matched to the
physical properties in the diffuse and translucent ISM at large.  The
lowest densities and interstellar radiation field strengths modeled in that
work are 1000~cm$^{-3}$ and 10 times the \citet{Draine78} field,
respectively, which are at the highest values in the \citet{Visser09} work.

In Figure \ref{cocivcoh2} we also overplot the cloud models of \citet{vDB88},
which included two values of the carbon depletion, $\delta_\mathrm{C}$.
The model with the higher value of depletion ($\delta_\mathrm{C}=0.1$) shows
a better agreement with the data than does the one with their preferred value
($\delta_\mathrm{C}=0.4$), which is consistent with observational
determinations of carbon depletion \citep{Sofia04}.

This discrepancy may be the result of multiple effects: systematic errors
in the oscillator strengths (e.g., use of the \citet{Morton03} values for
the \ci\ lines would produce a larger column densities); inappropriate
parameter space in the models, the results of which are sensitive to
densities, strengths of the interstellar radiation field, grain models,
chemical networks, etc.; or inaccuracies in the carbon depletion
determinations -- a recent study of the damping wings of the strong
\ion{C}{2} $\lambda 1335$ absorption line shows a lower gas phase abundance
than determinations from the weaker intersystem lines \citep{Sofia08}.

We note that \citet{Visser09} adopted the chemical network of
\citet{Rollig07}, which includes only species from the four most abundant
elements (H, He, O and C).  \citet{Rollig07} mentions that this network
excludes some species that could be important in the chemical balance of
carbon in photodissociation regions, such as sulfur.   Furthermore, the
chemical reactions governing the \ci\ abundance are complicated by the
presence of neutral and ionized polycyclic aromatic hydrocarbons
\citep{Bakes98,Wolfire08}, which can affect the position of the \ci\
abundance peak as a function of depth into a cloud.

Given that the models agree well with CO v H2, we can comment on their
ability to reproduce the CO/\htwo\ as a function of \av\ (see middle panel
of Figure 4).  The models show that the primary driver of CO/\htwo\ is
$\chi/n$, regardless of the extinction.  Our data span two orders of
magnitude change in $\chi/n$, with the highest CO columns appearing where
the $\chi/n$ is lowest.  For a given value of \av, geometric effects can
greatly change what the measured line-of-sight properties might be (see
Figure 4 of Kopp et al. 2000, for example) and thus extinction alone is not
a good discriminator of the presence of translucent material.

\section{Conclusions}

Based upon the results reported here, we suggest that the definition of
``translucent'' based on measures of extinction be replaced with one based
on molecular content, primarily CO.  In the range of \av\ sampled in this
study, extinction properties give no indication of the physical or chemical
conditions in the clouds sampled along the line of sight.  We believe this
to result primarily from the facts that extinction correlates better with
total hydrogen rather than molecular hydrogen and that the penetration of
ultraviolet radiation is very sensitive to cloud structure \citep{Kopp00}.

These observations provide good constraints for models of the interstellar
medium.  Significant column densities of CO, and high values of CO/\htwo,
even at relatively low extinction, suggest that the interstellar medium
cannot be too fragmented and porous; i.e., translucent material that is
sufficiently shielded from photodissociating far-UV radiation exists in
increasing abundance along sightlines with high molecular fraction.  These
sightlines cannot simply be concatenations of multiple diffuse clouds.

Finally, we point out that, like CO/\htwo, CO/\ci\ can be a good measure of
the presence of translucent material along a given line-of-sight. Although
the Cosmic Origins Spectrograph aboard the \textit{Hubble Space Telescope}
has some ability to measure \htwo\ along some sightlines
\citep{McCandliss09}, both it and the Space Telescope Imaging Spectrograph
are optimized for performance in the 1150--1550\AA\ range, where CO and
\ci\ both have many absorption features and may be measured simultaneously.

\acknowledgements We would like to thank R. Visser for providing us with
his model results.  Some/all of the data presented in this paper were
obtained from the Multimission Archive at the Space Telescope Science
Institute (MAST). STScI is operated by the Association of Universities for
Research in Astronomy, Inc., under NASA contract NAS5-26555. Support for
MAST for non-HST data is provided by the NASA Office of Space Science via
grant NAG5-7584 and by other grants and contracts.

\begin{deluxetable}{l||l|c|c|c||c|c||c|c|r||cc}
\tablewidth{6.5in}
\tabletypesize{\scriptsize} \tablecaption{Stellar Parameters, Extinctions, and Hydrogen Column Densities}
\tablehead{\colhead{Star Name} & \colhead{Sp. Type} & \colhead{E(B-V)} & \colhead{$A_V$} & \colhead{ref} & \colhead{Log N (HI)} & \colhead{ref} & \colhead{Log N (H2)} & \colhead{T$_{01}$} & \colhead{ref} & \colhead{$f^N$}}
\startdata
   CPD -59 2603  &                         07 V &  0.46 &   ... &  1 & 21.46 &  1 & 20.16$\pm$0.10 &   \phantom{1}77$\pm$             11 &  8 & 0.09$\pm$0.02\\
      HD 15137  &                       O9.5 V &  0.31 &   ... &  1 & 21.11 &  1 & 20.32$\pm$0.09 &             104$\pm$             21 &  9 & 0.24$\pm$0.05\\
      HD 24534  &                        B0 Ve &  0.59 &   ... &  1 & 20.73 &  1 & 20.92$\pm$0.04 &   \phantom{1}57$\pm$   \phantom{2}4 &  6 & 0.76$\pm$0.07\\
      HD 27778  &                         B3 V &  0.35 &  0.91 &  2 & 20.90 &  7 & 20.79$\pm$0.06 &   \phantom{1}55$\pm$   \phantom{2}7 &  6 & 0.61$\pm$0.08\\
      HD 37903  &                       B1.5 V &  0.35 &  1.31 &  2 & 21.17 &  1 & 20.92$\pm$0.06 &   \phantom{1}68$\pm$   \phantom{2}7 & 11 & 0.53$\pm$0.07\\
      HD 40893  &                        B1 IV &  0.47 &  1.32 &  2 & 21.50 &  5 & 20.58$\pm$0.05 &   \phantom{1}78$\pm$   \phantom{2}8 & 11 & 0.19$\pm$0.02\\
      HD 69106  &                      B0.5 II &  0.20 &  0.61 &  2 & 21.08 &  1 & 19.73$\pm$0.10 &   \phantom{1}80$\pm$             16 & 10 & 0.08$\pm$0.02\\
      HD 91824  &                         O6 V &  0.24 &  0.80 &  2 & 21.15 &  3 & 20.00$\pm$0.10 &   \phantom{1}61$\pm$   \phantom{2}7 &  8 & 0.12$\pm$0.03\\
      HD 93205  &                         O3 V &  0.38 &  1.23 &  2 & 21.33 &  1 & 19.86$\pm$0.10 &             105$\pm$             21 &  8 & 0.06$\pm$0.01\\
      HD 93222  &                         O8 V &  0.36 &  1.71 &  2 & 21.54 &  3 & 19.81$\pm$0.10 &   \phantom{1}77$\pm$             11 &  8 & 0.04$\pm$0.01\\
      HD 93843  &                       O6 III &  0.27 &  1.05 &  2 & 21.33 &  1 & 19.61$\pm$0.10 &             107$\pm$             21 & 10 & 0.04$\pm$0.01\\
      HD 99857  &                      B0.5 Ib &  0.33 &   ... &  1 & 21.31 &  1 & 20.25$\pm$0.10 &   \phantom{1}83$\pm$             17 & 10 & 0.15$\pm$0.03\\
      HD 102065  & B9 IV\tablenotemark{$\dagger$} &  0.17 &  0.67 & 13 & 20.49 &  7 & 20.53$\pm$0.10 &   \phantom{1}59$\pm$   \phantom{2}7 & 12 & 0.69$\pm$0.16\\
      HD 103779  &                      B0.5 II &  0.21 &  0.69 &  2 & 21.16 &  1 & 19.94$\pm$0.10 &   \phantom{1}86$\pm$             14 &  8 & 0.11$\pm$0.02\\
      HD 104705  &                       B0 III &  0.23 &  0.65 &  2 & 21.11 &  1 & 20.00$\pm$0.10 &   \phantom{1}92$\pm$             16 &  8 & 0.13$\pm$0.03\\
      HD 115071  &                       B0.5 V &  0.49 &   ... &  1 & 21.38 &  1 & 20.69$\pm$0.09 &   \phantom{1}71$\pm$             14 &  9 & 0.29$\pm$0.06\\
      HD 116852  &                       O9 III &  0.21 &  0.51 &  2 & 20.96 &  1 & 19.86$\pm$0.10 &   \phantom{1}70$\pm$   \phantom{2}9 &  8 & 0.14$\pm$0.03\\
      HD 121968  &                         B1 V &  0.07 &   ... &  1 & 20.71 &  1 & 18.73$\pm$0.10 &   \phantom{1}38$\pm$   \phantom{2}3 &  8 & 0.02$\pm$0.00\\
      HD 124314  &                         O6 V &  0.53 &   ... &  1 & 21.34 &  1 & 20.52$\pm$0.09 &   \phantom{1}74$\pm$             15 &  9 & 0.23$\pm$0.05\\
      HD 147888  &                         B4 V &  0.51 &  1.99 &  2 & 21.34 &  7 & 20.58$\pm$0.09 &   \phantom{1}44$\pm$   \phantom{2}9 &  9 & 0.26$\pm$0.05\\
      HD 152590  &                       O7.5 V &  0.38 &   ... &  4 & 21.37 &  4 & 20.51$\pm$0.09 &   \phantom{1}64$\pm$             13 &  9 & 0.22$\pm$0.04\\
      HD 157857  &                         O7 V &  0.43 &  1.48 &  2 & 21.30 &  1 & 20.69$\pm$0.09 &   \phantom{1}86$\pm$             17 &  9 & 0.33$\pm$0.07\\
      HD 177989  &                        B2 II &  0.23 &  0.65 &  2 & 20.95 &  1 & 20.24$\pm$0.10 &   \phantom{1}52$\pm$   \phantom{2}5 &  8 & 0.28$\pm$0.06\\
      HD 185418  &                       B0.5 V &  0.50 &  1.27 &  2 & 21.11 &  3 & 20.80$\pm$0.10 &             105$\pm$             21 &  8 & 0.49$\pm$0.11\\
      HD 192639  &                         O8 V &  0.61 &  1.91 &  2 & 21.32 &  1 & 20.75$\pm$0.09 &   \phantom{1}98$\pm$             15 &  9 & 0.35$\pm$0.07\\
      HD 201345  &                         O9 V &  0.18 &   ... &  1 & 20.88 &  1 & 19.46$\pm$0.10 &             147$\pm$             41 &  8 & 0.07$\pm$0.02\\
      HD 203374  &                        B0 IV &  0.22 &   ... &  1 & 21.11 &  1 & 20.68$\pm$0.10 &   \phantom{1}87$\pm$             17 & 10 & 0.43$\pm$0.10\\
      HD 203532  &                         B5 V &  0.28 &  0.94 &  2 & 20.78 &  7 & 20.71$\pm$0.10 &   \phantom{1}49$\pm$   \phantom{2}5 &  8 & 0.63$\pm$0.15\\
      HD 206267  &                         O6 V &  0.52 &  1.47 &  2 & 21.30 &  6 & 20.86$\pm$0.04 &   \phantom{1}65$\pm$   \phantom{2}5 &  6 & 0.42$\pm$0.04\\
      HD 207198  &                        O9 II &  0.54 &  1.50 &  2 & 21.34 &  1 & 20.83$\pm$0.04 &   \phantom{1}66$\pm$   \phantom{2}5 &  6 & 0.38$\pm$0.04\\
      HD 210839  &                       O6 Iab &  0.57 &  1.15 &  2 & 21.15 &  1 & 20.84$\pm$0.04 &   \phantom{1}72$\pm$   \phantom{2}6 &  6 & 0.49$\pm$0.05\\
      HD 218915  &                     09.5 Iab &  0.29 &   ... &  1 & 21.11 &  1 & 20.16$\pm$0.10 &   \phantom{1}86$\pm$             14 &  8 & 0.18$\pm$0.04\\
      HD 224151  &                      B0.5 II &  0.44 &   ... &  1 & 21.32 &  1 & 20.57$\pm$0.10 &             252$\pm$             50 & 10 & 0.26$\pm$0.06\\
      HD 303308  &                         O3 V &  0.45 &  1.36 &  2 & 21.45 &  1 & 20.36$\pm$0.10 &   \phantom{1}86$\pm$             14 &  8 & 0.14$\pm$0.03\\
\enddata
\label{table:hydrogen}
\tablenotetext{$\dagger$}{See text for a discussion of the spectral type of HD~102065}
\tablerefs{(1) \citet{Diplas94}, (2) \citet{Valencic04}, (3) \citet{FM90}, (4) \citet{Cartledge04}, (5) \citet{Jensen07}, (6) \citet{Rachford02}, (7) This paper; N(\ion{H}{1}) = $5.8\times10^{21}E(B-V)-2N$(\htwo) as per \citet{Bohlin78}, (8) \citet{Burgh07}, (9) \citet{Sheffer08}, (10) \citet{Shull09}, (11) \citet{Rachford09}, (12) \citet{Gry02}, (13) \citet{Nehme08a} }
\end{deluxetable}

\begin{deluxetable}{l||l|c||c||rr}
\tablewidth{6in}
\tabletypesize{\scriptsize} \tablecaption{Carbon Column Densities}
\tablehead{\colhead{Star Name} & \colhead{Log N (CO)} & \colhead{ref} & \colhead{Log N (\ion{C}{1})\tablenotemark{$\dagger$}} & \colhead{CO/\ion{C}{1}} & \colhead{err}}
\startdata
   CPD -59 2603  &  \phantom{<<<}14.15 $\pm$  0.10 &   1 & 14.65 &     0.32 &     0.08\\
      HD 15137  &  \phantom{<<<}13.52 $\pm$  0.09 &   3 & 14.60 &     0.08 &     0.02\\
      HD 24534  &  \phantom{<<<}16.13 $\pm$  0.20 &   1 & 14.98 &    14.08 &     7.25\\
      HD 27778  &  \phantom{<<<}16.05 $\pm$  0.14 &   1 & 15.06 &     9.79 &     3.35\\
      HD 37903  &                  $<$13.70       &   2 & 14.22 &  $<$0.30 &         \\
      HD 40893  &  \phantom{<<<}14.18 $\pm$  0.20 &   2 & 14.65 &     0.34 &     0.16\\
      HD 69106  &                  $<$13.50       &   2 & 14.27 &  $<$0.17 &         \\
      HD 91824  &                  $<$13.60       &   1 & 14.47 &  $<$0.14 &         \\
      HD 93205  &  \phantom{<<<}13.23 $\pm$  0.06 &   1 & 14.54 &     0.05 &     0.01\\
      HD 93222  &  \phantom{<<<}13.36 $\pm$  0.20 &   1 & 14.36 &     0.10 &     0.05\\
      HD 93843  &                  $<$12.70       &   2 & 14.13 &  $<$0.04 &         \\
      HD 99857  &  \phantom{<<<}13.73 $\pm$  0.10 &   2 & 14.59 &     0.14 &     0.04\\
      HD 102065  &  \phantom{<<<}13.62 $\pm$  0.12 &   1 & 14.24 &     0.24 &     0.07\\
      HD 103779  &                  $<$12.35       &   1 & 14.21 &  $<$0.01 &         \\
      HD 104705  &  \phantom{<<<}12.98 $\pm$  0.16 &   1 & 14.24 &     0.05 &     0.02\\
      HD 115071  &  \phantom{<<<}14.53 $\pm$  0.09 &   3 & 14.69 &     0.69 &     0.16\\
      HD 116852  &  \phantom{<<<}13.28 $\pm$  0.04 &   1 & 14.15 &     0.13 &     0.02\\
      HD 121968  &                  $<$12.30       &   1 & 13.36 &  $<$0.09 &         \\
      HD 124314  &  \phantom{<<<}14.20 $\pm$  0.09 &   3 & 14.67 &     0.34 &     0.08\\
      HD 147888  &  \phantom{<<<}15.28 $\pm$  0.09 &   3 & 14.70 &     3.80 &     1.17\\
      HD 152590  &  \phantom{<<<}13.77 $\pm$  0.09 &   3 & 14.60 &     0.15 &     0.03\\
      HD 157857  &  \phantom{<<<}14.08 $\pm$  0.09 &   3 & 14.62 &     0.29 &     0.07\\
      HD 177989  &  \phantom{<<<}14.62 $\pm$  0.17 &   1 & 14.68 &     0.87 &     0.35\\
      HD 185418  &  \phantom{<<<}14.82 $\pm$  0.20 &   1 & 14.74 &     1.20 &     0.57\\
      HD 192639  &  \phantom{<<<}13.78 $\pm$  0.09 &   3 & 14.73 &     0.11 &     0.03\\
      HD 201345  &                  $<$12.40       &   1 & 13.93 &  $<$0.03 &         \\
      HD 203374  &  \phantom{<<<}15.35 $\pm$  0.10 &   2 & 14.98 &     2.34 &     0.60\\
      HD 203532  &  \phantom{<<<}15.70 $\pm$  0.17 &   1 & 14.75 &     8.95 &     4.07\\
      HD 206267  &  \phantom{<<<}16.11 $\pm$  0.13 &   1 & 15.32 &     6.10 &     1.96\\
      HD 207198  &  \phantom{<<<}15.53 $\pm$  0.20 &   1 & 15.26 &     1.88 &     0.89\\
      HD 210839  &  \phantom{<<<}15.41 $\pm$  0.08 &   1 & 15.00 &     2.59 &     0.56\\
      HD 218915  &  \phantom{<<<}13.64 $\pm$  0.13 &   1 & 14.54 &     0.13 &     0.04\\
      HD 224151  &  \phantom{<<<}13.85 $\pm$  0.06 &   2 & 14.63 &     0.17 &     0.03\\
      HD 303308  &  \phantom{<<<}13.65 $\pm$  0.06 &   1 & 14.73 &     0.08 &     0.02\\
\enddata
\label{table:carbon}
\tablenotetext{$\dagger$}{Errors for \ion{C}{1} are assumed to be 0.05 dex in
  all cases except for HD~24534, HD~147888, and HD~203534, which are
  assigned an error of 0.10 dex.  See text for discussion.}
\tablerefs{(1) \citet{Burgh07}, (2) This paper, (3) \citet{Sheffer08}}
\end{deluxetable}

\end{document}